\documentclass[a4paper,11pt]{article}
\pdfoutput=1
\usepackage{amssymb,amsmath,amsfonts,makeidx,placeins,pbox,multirow,mathtools,graphicx,rotate,subcaption,color,slashed,cite,caption,epstopdf,verbatim,url,booktabs,longtable,tabu,bm,array,mathrsfs}
\newcolumntype{P}[1]{>{\centering\arraybackslash}p{#1}}
\usepackage[colorlinks=true,
            linkcolor=magenta,
            urlcolor=blue,
            citecolor=blue]{hyperref}
\usepackage[utf8x]{inputenc}
\usepackage{siunitx}
\usepackage[capitalise]{cleveref}
\usepackage[normalem]{ulem}

\crefname{section}{Sec.}{Secs.}
\crefname{table}{Tab.}{Tabs.}
\crefname{figure}{Fig.}{Figs.}
\crefname{equation}{Eq.}{Eqs.}
\crefname{appendix}{Appendix}{Appendix}

\long\def\rpl#1!!#2!!{\textcolor{red}{\sout{#1}} \textcolor{blue}{#2}}
\long\def\rpll#1!!#2!!{\textcolor{red}{#1} \textcolor{blue}{#2}}
\newcommand{\tQ}{t_3^{(+Q)}}
\newcommand{\tone}{t_3^{(+1)}}
\newcommand{\ttwo}{t_3^{(+2)}}
\def\order(#1){{\cal O} \left(#1 \right)}

\evensidemargin=\oddsidemargin
\begin{document}

\begin{flushright}
\begin{small}
TIFR/TH/25-2
\end{small}
\end{flushright}
\vspace{0.2cm}

\begin{center}

{\Large \bf Drell-Yan constraints on charged scalars: \\ a weak isospin perspective} \\

\vspace*{0.5cm} {\sf Avik
  Banerjee~$^{a,}$\footnote{avik.banerjee\_205@tifr.res.in}, ~Dipankar Das~$^{b,}$\footnote{d.das@iiti.ac.in}, ~Samadrita Mukherjee~$^{c,}$\footnote{samadritam@iisc.ac.in}, ~Shreya Pandey~$^{b,}$\footnote{shreyavatspandey@gmail.com}} \\
\vspace{10pt} {\small } $^{a)}$ {\em Department of Theoretical Physics, Tata Institute of Fundamental Research,\\ Homi Bhabha Road, Mumbai 400005, India}
\\
\vspace{5pt} {\small } $^{b)}$ {\em Indian Institute of Technology (Indore), Khandwa Road, Simrol, Indore 453 552, India} 
\\
\vspace{5pt} {\small } $^{c)}$ {\em Centre for High Energy Physics, Indian Institute of Science, Bangalore 560012, India} \normalsize
\end{center}


\begin{abstract}
Charged scalars appear in many motivated extensions beyond the Standard Model. We analyze the constraints on charged scalar pair production via the Drell-Yan process at the Large Hadron Collider and interpret them in terms of weak isospin quantum numbers. Leveraging the experimental limits from existing LHC data and phenomenological recast analyses, we place bounds on the branching ratio of the charged scalar, as a function of its mass, electric charge, and isospin. This approach enables to determine limits on the branching ratios directly from experimental data, without appealing to a specific model.  We provide a detailed analysis for singly and doubly charged scalars across various weak isospin scenarios, focusing on decays into leptonic and bosonic final states, and validate this approach in extended Higgs sectors such as the Higgs triplet model and Georgi-Machacek model.
\end{abstract}


\bigskip

\begin{figure}[t!]
	\centering
	\includegraphics[width=0.4\linewidth]{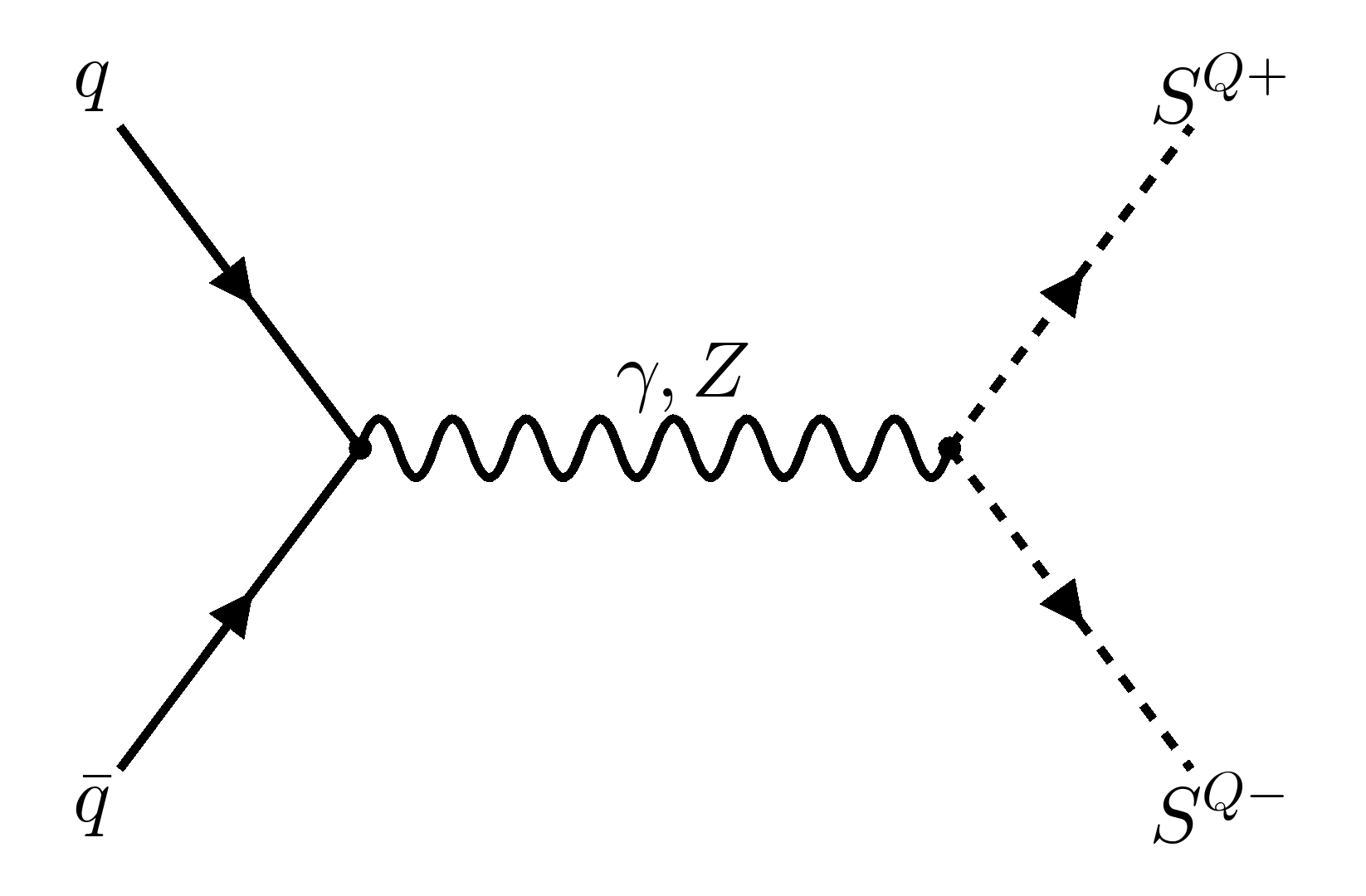}
	\caption{\em Feynman diagram contributing to the DY pair production of charged scalars at hadron collider. Here, the arrows denote the direction of the momentum.}
	\label{fig:pair_production}
\end{figure}

\paragraph{Introduction:}
\label{intro}
Numerous motivated scenarios beyond the Standard Model (BSM) propose the existence of novel scalar particles with non-zero electric charges. Pair production via the Drell Yan (DY) process (illustrated in \cref{fig:pair_production} through the corresponding Feynman diagram) at the Large Hadron Collider (LHC) represents a key discovery channel for such new electrically charged particles. Notably, DY production is routinely employed by the ATLAS and CMS collaborations to search for doubly charged scalars. Doubly charged scalars appear in many BSM scenarios and are favored for their enhanced production cross-section, which arises from their relatively higher electric charge. As a result, doubly charged scalars are a primary focus of DY searches at the LHC~\cite{Crivellin:2018ahj,Fuks:2019clu}. So far no indications of BSM physics have been observed at the LHC, and the results from these searches are usually presented as upper limits on cross-sections. In the current context, the bounds obtained from the DY searches are typically reinterpreted in a model-specific way by the experimental collaborations\cite{ALEPH:2002ftu,DELPHI:2003eid,OPAL:2008tdn,L3:2003jyb,ALEPH:2013htx}.

This work stems from the observation that the DY pair production cross-section of charged scalars at the LHC is solely dependent on their masses, electric charge and isospin quantum numbers \cite{delAguila:2013mia}. This general feature of the production cross-section can be leveraged to derive constraints on the branching ratios of charged scalars in a largely model-independent way, using current experimental data from the LHC \cite{ATLAS:2022pbd,ATLAS:2021jol} and recent results from phenomenological recast analyses \cite{Banerjee:2022xmu,Cacciapaglia:2022bax}. 

To make the discussion explicit, let us denote a colorless scalar field with electric charge $\pm Q$ as $S^{Q\pm}$. The Lagrangian governing the interactions of these scalar fields with a $Z$ boson and a photon can be written in a compact form as\footnote{In passing we mention that the $W$ mediated production process becomes relevant in the presence of multiple scalars with electric charge difference $|\Delta Q|=1$. Additionally, it involves extra parameters, such as the total isospin of the scalar multiplet as well as the mass difference between the two scalars. We focus only on the $Z$ and photon-mediated processes in this article, since the current searches at the LHC have rather weak sensitivity to the $W$-mediated production channels.}
\begin{align}
{\mathscr L} &= ie \left( Q A^\mu + \frac{K^{Q}_{Z}}{s_Wc_W} Z^\mu \right) \bigg[(\partial_\mu S^{Q+}) S^{Q-} - S^{Q+}(\partial_\mu S^{Q-}) \bigg],
\label{L_SSV} 
\end{align}
where $s_W (c_W)\equiv \sin\theta_W (\cos\theta_W)$ denotes the weak mixing, and $e$ is the electromagnetic coupling constant. The coefficient $K^{Q}_{Z}$ can be expressed purely in terms of the isospin and the electric charge of the scalars as\cite{delAguila:2013mia}
\begin{eqnarray}
    K^Q_{Z} = \tQ - Q s_W^2,
    \label{coeffs}
\end{eqnarray}
where $\tQ$ denotes the eigenvalue of $S^{Q+}$ along the diagonal generator $T_{3L}$ of the $SU(2)_L$. In our notation, the hypercharge of the multiplet is related to the electric charge as $Q=T_{3L}+Y$. 

Clearly, the production cross-section for a $S^{Q\pm}$ pair via DY process, mediated by a photon or $Z$ as shown in \cref{fig:pair_production}, depends on the electric charge $Q$, and the third component of the isospin $t^{(+Q)}_3$, in addition to the scalar mass $m_{S^{Q+}}$.\footnote{It is important to realize that the DY pair production cross-section depends not on the $SU(2)_L$ multiplet itself, but on the specific position of the charged scalar within that multiplet.} Consequently, there is no difference between a doubly charged scalar arising from the  Higgs Triplet Model~(HTM) \cite{Konetschny:1977bn,Cheng:1980qt,Schechter:1980gr,Dey:2008jm} and one from the Georgi-Machacek~(GM) model~\cite{Georgi:1985nv}, as long as the DY pair-production is concerned. In the same way, the right handed doubly charged scalar in the Left-Right Symmetric Model (LRSM)~\cite{Mohapatra:1979ia, Deshpande:1990ip} will have the same DY production cross-section as the doubly charged scalar from the Zee-Babu model~\cite{Zee:1985id,Babu:1988ki}. Our main point, therefore, is that the results from the DY searches can be concisely classified in terms of $Q$ and $\tQ$ values of the charged scalars. This approach allows for the extraction of model-specific information without the need to simulate each model individually, thereby optimizing the use of computational resources.

In this article, we aim to provide a convenient and easy to interpret catalog of constraints utilizing the current experimental data from the DY searches for singly and doubly charged scalars at the LHC, along with the recent results from phenomenological recast analyses. We specifically focus on the singly and doubly charged scalars with varying $\tQ$. It is important to note that, unlike the doubly charged scalars, no DY-based search results for singly charged scalars are currently available from the LHC, presumably due to their relatively small production cross-section. Therefore, for singly-charged scalars, we will rely on the projections from phenomenological analyses, such as those in \cite{Banerjee:2022xmu, Cacciapaglia:2022bax}.

\paragraph{Summary of extended Higgs sector models:}
\label{models}
Before presenting our main results, we provide a brief summary of the key aspects of the popular models covered in our analysis. This overview will also help clarify Table~\ref{tab:my_label}, which categorizes the BSM charged scalars from several well-motivated models based on their electric charges and $\tQ$ values.

\begin{itemize}
\item {\em Multi Higgs-doublet models}~\cite{Grossman:1994jb} contain singly charged scalars with $\tone = +\frac{1}{2}$. Note that, even when the physical charged scalars arise from the mixing among two or more unphysical component fields which have identical $\tQ$ eigenvalues, the final DY production cross-section remains unaffected by the mixing.
\item {\em The HTM}~\cite{Konetschny:1977bn,Cheng:1980qt,Schechter:1980gr,Dey:2008jm} introduces a complex $SU(2)_L$ triplet with $Y=1$ alongside the SM scalar
doublet. This model includes a doubly charged scalar with $\ttwo= +1$ as well as a singly charged scalar arising from the mixing between the components of the doublet and the triplet. However, for practical purposes, in the limit the vacuum expectation value (VEV) of the triplet is much smaller than the doublet VEV~\cite{Melfo:2011nx,Arhrib:2011uy,Kanemura:2012rs,Das:2016bir}, the singly charged scalar can be treated as having $\tone=0$.
Similarly, for the SM extended by a real Higgs-triplet~($Y=0$)~\cite{Butterworth:2023rnw,Ashanujjaman:2024lnr}, there will be a singly charged scalar with $\tone=+1$.

\item {\em The Georgi-Machacek~(GM) model}~\cite{Georgi:1985nv, Chanowitz:1985ug} extends the scalar sector of the SM by adding one real ($Y=0$) and one complex ($Y=1$) Higgs-triplets. The doubly charged scalar purely stems from the $\ttwo= +1$ component field of the complex triplet. On the other hand, the physical singly charged
scalars arise from a mixing between the $\tone=0$ and $\tone= +1$ component fields with well defined mixing angle~\cite{Gunion:1989ci}.\footnote{Strictly speaking, for the electrically charged member of the custodial triplet, there should be additional mixing from the $\tQ=+\frac{1}{2}$ component of the doublet as well. However, in the limit when the doublet VEV is overwhelmingly dominant over the common triplet VEV~\cite{Chakraborti:2023mya}, we can ignore this mixing.}
\item {\em The Zee-type model}~\cite{Zee:1980ai} relies on a two Higgs-doublet structure~\cite{Branco:2011iw,Bhattacharyya:2015nca} 
augmented by an $SU(2)_L$ scalar singlet with $Y=1$. The model leads to two physical singly charged scalars arising from the mixing between $\tone=0$ and $\tone=+\frac{1}{2}$ components~\cite{Florentino:2021ybj}. This mixing angle, usually, is a free parameter.
\item {\em The Zee-Babu model}~\cite{Zee:1985id,Babu:1988ki} extends the scalar sector of the SM by adding two complex scalar
$SU(2)_L$ singlets with $Y=1$ and $Y=2$ respectively. As a result, the physical singly charged and
doubly charged scalars both have $\tQ=0$~\cite{Herrero-Garcia:2014hfa}. The DY production phenomenology will be very similar to other
models, which also extend the scalar sector of the SM with complex $SU(2)_L$ singlets with nonzero
hypercharges~\cite{Barger:2008jx, Boos:2018fnt}.
\item The scalar sector of the {\em conventional left-right symmetric model~(LRSM)}~\cite{Mohapatra:1979ia, Deshpande:1990ip} consists of a bidoublet,
an $SU(2)_L$ triplet and an $SU(2)_R$ triplet.\footnote{The initial versions of the model~\cite{Mohapatra:1974gc, Senjanovic:1975rk} contain,
in addition to the bidoublet, two doublets of $SU(2)_L$ and $SU(2)_R$ respectively. Consequently,
doubly charged scalars do not arise from such a model.} The VEV of the $SU(2)_L$ triplet is much smaller than
the bidoublet VEVs which, in turn, are much smaller than the VEV of the $SU(2)_R$ triplet. For such
hierarchies among the VEVs~\cite{Duka:1999uc}, the two physical singly charged scalars have $\tone=0$ and $\tone=+\frac{1}{2}$ respectively. Furthermore, the two physical doubly charged scalars have $\ttwo=+1$ and $\ttwo=0$ respectively.
\item {\em The two Higgs doublet model (2HDM) where the SM scalar sector is extended by an $SU(2)_L$ scalar doublet with $Y=\frac{3}{2}$} contains an unusual doubly charged scalar with $\ttwo =+\frac{1}{2}$\cite{Enomoto:2021fql}. The physical singly charged scalar in this model arises purely from the $\tone=-\frac{1}{2}$ component of the additional doublet.
\item {\em The Higgs-Septet model} contains a plethora of exotic charged scalars~\cite{Hisano:2013sn}. Among them, the doubly charged scalar in this model arises from the $\ttwo=0$ component of the septet. The singly charged scalars, under the assumption that the septet VEV is much smaller than the doublet VEV~\cite{Alvarado:2014jva}, arises from a mixing between the $\tone=-1$ and $\tone=+3$ components of the septet~\cite{Harris:2017ecz}.

\item {\em The composite Higgs models}~based on the $SU(5)/SO(5)$ \cite{Dugan:1984hq,Ferretti:2014qta} and $SU(4)\times SU(4)^{\prime}/SU(4)_V$ \cite{Ma:2015gra,Wu:2017iji} cosets feature doubly and singly charged scalars as pseudo Nambu-Goldstone bosons (pNGBs). The former coset gives rise to a singlet ($Y=0$), a real ($Y=0$) and a complex triplet ($Y=1$) in addition to the usual Higgs doublet. This is quite similar to the GM model, however, the triplets have opposite CP properties compared to the GM model, and they do not receive any VEVs. The doubly charged scalar arises from the $\ttwo=+1$ component of the complex triplet. In contrast, the physical singly charged scalars are admixture of $\tone=+1$ component of the real and $\tone=0$ component of the complex triplet. The $SU(4)\times SU(4)^\prime/SU(4)_V$ coset gives rise to fifteen pNGBs including two doublets (with $Y=\frac{1}{2}$), one real triplet ($Y=0$) and three singlets (one with $Y=1$ and other two with $Y=0$). Here three singly charged scalars arise from the $\tone=+\frac{1}{2}$, $\tone=+1$ and $\tone=0$ components of the doublet, triplet and the singlet ($Y=1$), respectively.

\end{itemize}
\begin{table}[t!]
		\setlength{\tabcolsep}{10pt} 
		\renewcommand{\arraystretch}{1.5} 
		\begin{center}
            \resizebox{\textwidth}{!}{%
			\begin{tabular}{c|c|c|c} 
				\toprule
				Electric Charge  & $t_3$ Eigenvalues  & $SU(2)_L$ Multiplet & Models \\
				(Q) & $\left(\tQ\right)$ & (t) & \\
				\midrule
				
				& 0& Singlet (t=0) & Zee-Babu Model~\cite{Zee:1985id,Babu:1988ki}, LRSM \cite{Mohapatra:1979ia, Deshpande:1990ip}\\ 
				\cmidrule{3-4}
				& & Septet (t=3) & Higgs-Septet Model~\cite{Hisano:2013sn} \\
				\cmidrule{2-4}
				Doubly Charged & +$\frac{1}{2}$ & Doublet (t$=\frac{1}{2}$)& SM + Doublet with $Y=\frac{3}{2}$ \cite{Enomoto:2021fql} \\ 
				\cmidrule{2-4}
				$(Q=+2)$ & +1 & Triplet (t=1) & HTM~\cite{Konetschny:1977bn,Cheng:1980qt,Schechter:1980gr} \\
				&    &               & GM Model~\cite{Georgi:1985nv,Chanowitz:1985ug}, LRSM \cite{Mohapatra:1979ia, Deshpande:1990ip}\\
				\midrule	
				& 0  & Singlet(t=0) & Zee-Babu Model~\cite{Zee:1985id,Babu:1988ki}, LRSM \cite{Mohapatra:1979ia, Deshpande:1990ip} \\
				\cmidrule{3-4}
				&    & Triplet (t=1) & HTM (Complex)~\cite{Dey:2008jm,Arhrib:2011uy,Kanemura:2012rs,Das:2016bir}, LRSM \cite{Mohapatra:1979ia, Deshpande:1990ip} \\
				\cmidrule{2-4}
				   & +$\frac{1}{2}$ & Doublet (t$=\frac{1}{2}$) & 2HDMs, Multi HDMs~\cite{Branco:2011iw,Bhattacharyya:2015nca, Florentino:2021ybj, Enomoto:2021fql},\\
				& & & LRSM \cite{Mohapatra:1979ia, Deshpande:1990ip} \\
                \cmidrule{2-4}
				Singly Charged & -$\frac{1}{2}$ & Doublet (t$=\frac{1}{2}$) & SM + Doublet with $Y=\frac{3}{2}$ \cite{Enomoto:2021fql} \\
				\cmidrule{2-4}
				$(Q=+1)$ &  +1 & Triplet (t=1) & Real Triplet~\cite{Butterworth:2023rnw,Ashanujjaman:2024lnr}\\
				\cmidrule{2-4}
                    & Mixed & Septet (t=3) & Higgs-Septet Model \cite{Hisano:2013sn} \\
				& ($-1$ and +$3$) & & \\
                    \cmidrule{2-4}
                    & Mixed & Singlet and & Zee model \cite{Zee:1980ai}\\
				& (0 and +$\frac{1}{2}$) & Doublet & \\
                    \cmidrule{2-4}
				& Mixed & Real and & GM Model \cite{Georgi:1985nv,Chanowitz:1985ug}\\
				& (0 and +1) & Complex Triplets & \\
				\bottomrule
			\end{tabular}}	
		\end{center}
		\caption{\sf\it Categorization of BSM scenarios featuring singly ($Q=1$) and doubly ($Q=2$) charged scalars with different $\tQ$ values.}
		\label{tab:my_label}
\end{table}

\paragraph{Methodology and results:}
\label{BR_limits}

 As declared earlier in this paper, we focus on singly ($Q=+1$) and doubly ($Q=+2$) charged scalars only and assume that they have narrow widths. We emphasize again that the pair production cross-section depends on $Q$, $\tQ$ and the masses of the scalars.

Consider a specific process $pp \to S^{Q+}S^{Q-} \to F_1 \bar F_2$, where $F_1$ and $\bar F_2$ denote the final states resulting from the decays of $S^{Q+}$ and $S^{Q-}$, respectively. For example, in the case of $S^{++}\to e^+ e^+$ decay, $F_1\equiv e^+ e^+$. The experimental collaborations and the phenomenological recasts usually place exclusion limits on the production cross-section times branching ratios as a function of the charged scalar mass at 95\% CL, denoted by $\sigma_{95}(m_{S^{Q+}})$. Following a simplified approach, these bounds can then be reinterpreted as constraints on the branching ratios, as follows:
\begin{align}
    BR\left(S^{Q+}\to F_1\right) BR\left(S^{Q-}\to \bar{F}_2\right) \leq \frac{\sigma_{95}(m_{S^{Q+}})}{\sigma_{\rm prod}(m_{S^{Q+}})}\,,
    \label{eq:bounds}
\end{align}
where $\sigma_{\rm prod}(m_{S^{Q+}})$ is the DY production cross-section, which depends on $Q$, $\tQ$ and $m_{S^{Q+}}$. For $F_1=F_2$, Eq.\,\eqref{eq:bounds} reduces to~\cite{Ruiz:2022sct}
\begin{align}
    BR\left(S^{Q+}\to F_1\right)\leq \sqrt{\frac{\sigma_{95}(m_{S^{Q+}})}{\sigma_{\rm prod}(m_{S^{Q+}})}}\,.
    \label{eq:bounds_FF}
\end{align}
We have implemented the relevant interactions in \texttt{FeynRules} \cite{Alloul:2013bka} and, generated a \texttt{UFO} \cite{Degrande:2011ua} model file 
using \texttt{FeynArts} \cite{Hahn:2000kx} and \texttt{NLOCT} \cite{Degrande:2014vpa}, and obtained the DY cross-section $\sigma_{\rm prod}(m_{S^{Q+}})$ at NLO in QCD \cite{Muhlleitner:2003me} using the \texttt{MG5\_aMC}~\cite{Alwall:2014hca}.
In this way, the above expressions provide a straightforward method for interpreting experimental bounds as constraints on the branching ratios, relying solely on the electric charge and isospin quantum numbers of the charged scalar, without invoking any specific model. Notably, the LEP analysis focuses exclusively on the 2HDM, where the charged Higgs originates from the $\tone=+\frac{1}{2}$ component of the doublet. A key result of our analysis is that it illustrates how the same methodology can be extended to accommodate general $\tone$ values.

\paragraph{Doubly charged scalars:}
\label{sec:limit_SQ2}

First, we will discuss the DY pair-production of doubly charged scalars. The doubly-charged scalars are usually looked for in the following decay channels: 
$$
S^{++}\to \ell^+\ell^{\prime +}, W^+ W^+\,, \quad {\rm where} \quad \ell, \ell' \equiv e, \mu\,.
$$
The doubly charged scalar in the composite Higgs model based on the $SU(5)/SO(5)$ coset undergoes a three-body decay $S^{++}\to W^+ t\bar b$ with nearly 100\% $BR$ via an off-shell $S^+$ exchange, for more details see~\cite{Banerjee:2022xmu, Cacciapaglia:2022bax, Flacke:2023eil, Banerjee:2023upj}. However, the current limits\cite{Cacciapaglia:2022bax} are not strong enough to impose bounds on this channel for $|\ttwo|\leq 1$. Thus, we only focus on the two-body decays of $S^{++}$ in this article. 

\begin{figure}[t!]
    \centering
    \includegraphics[width=0.42\textwidth,height=0.20\textheight]{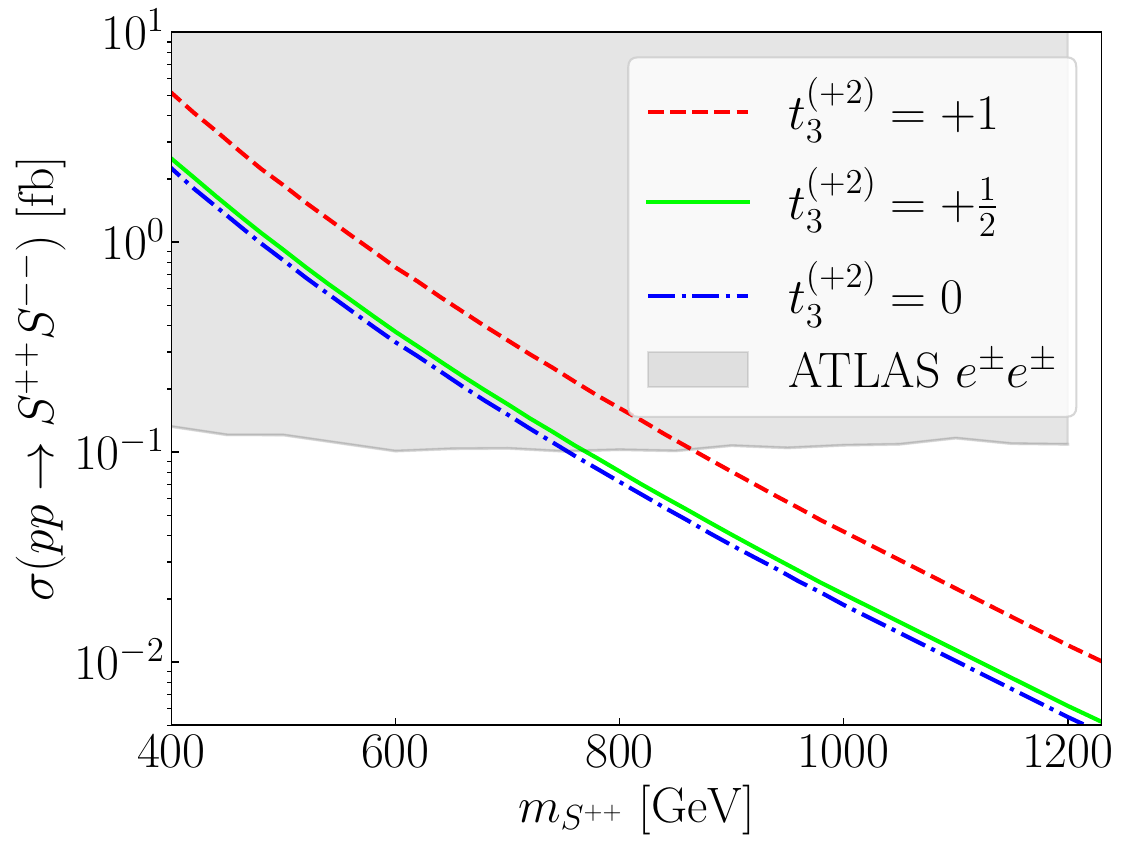}
    \includegraphics[width=0.42\textwidth,height=0.20\textheight]{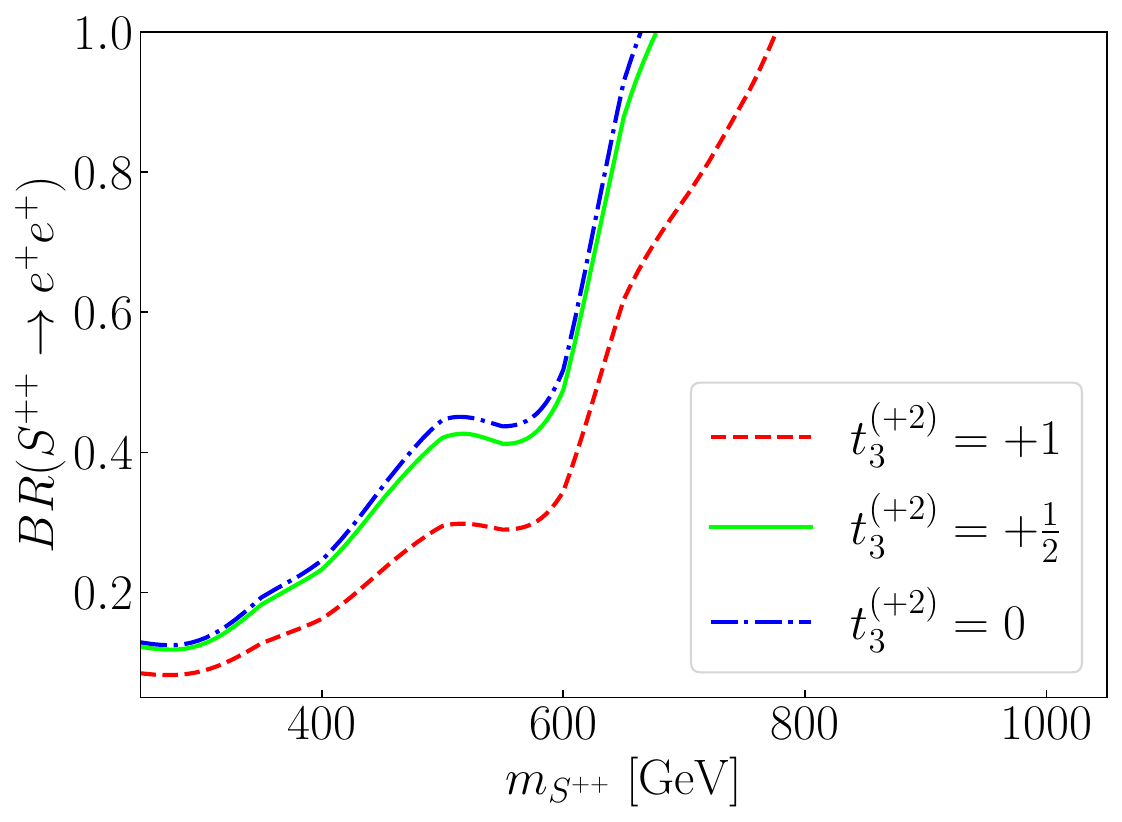}
    \includegraphics[width=0.42\textwidth,height=0.20\textheight]{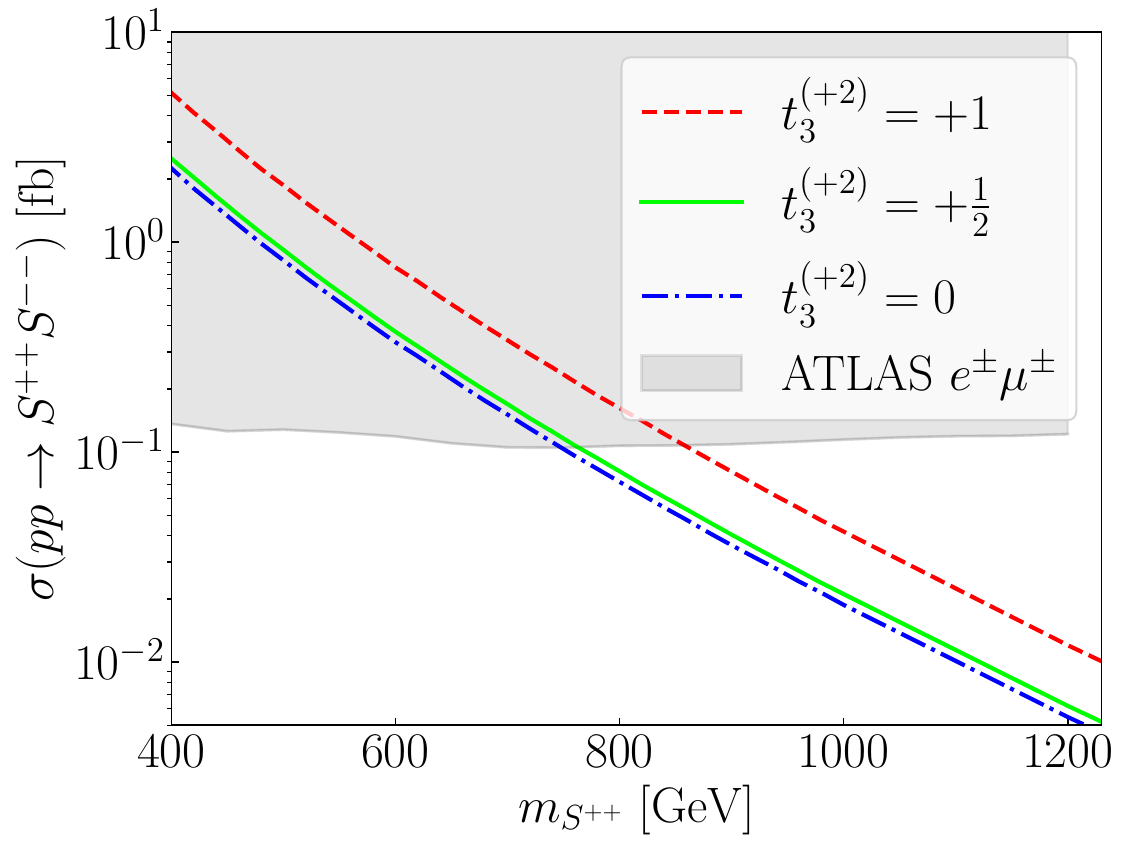}
    \includegraphics[width=0.42\textwidth,height=0.20\textheight]{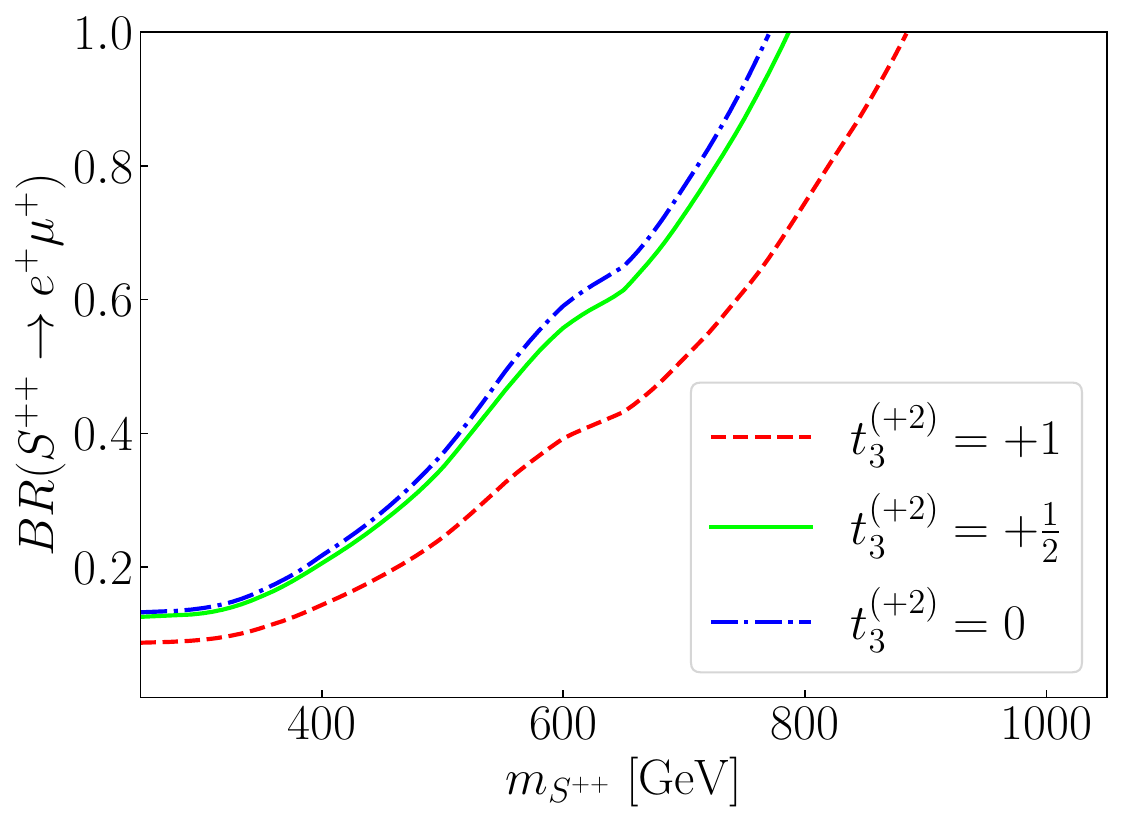}
    \includegraphics[width=0.42\textwidth,height=0.20\textheight]{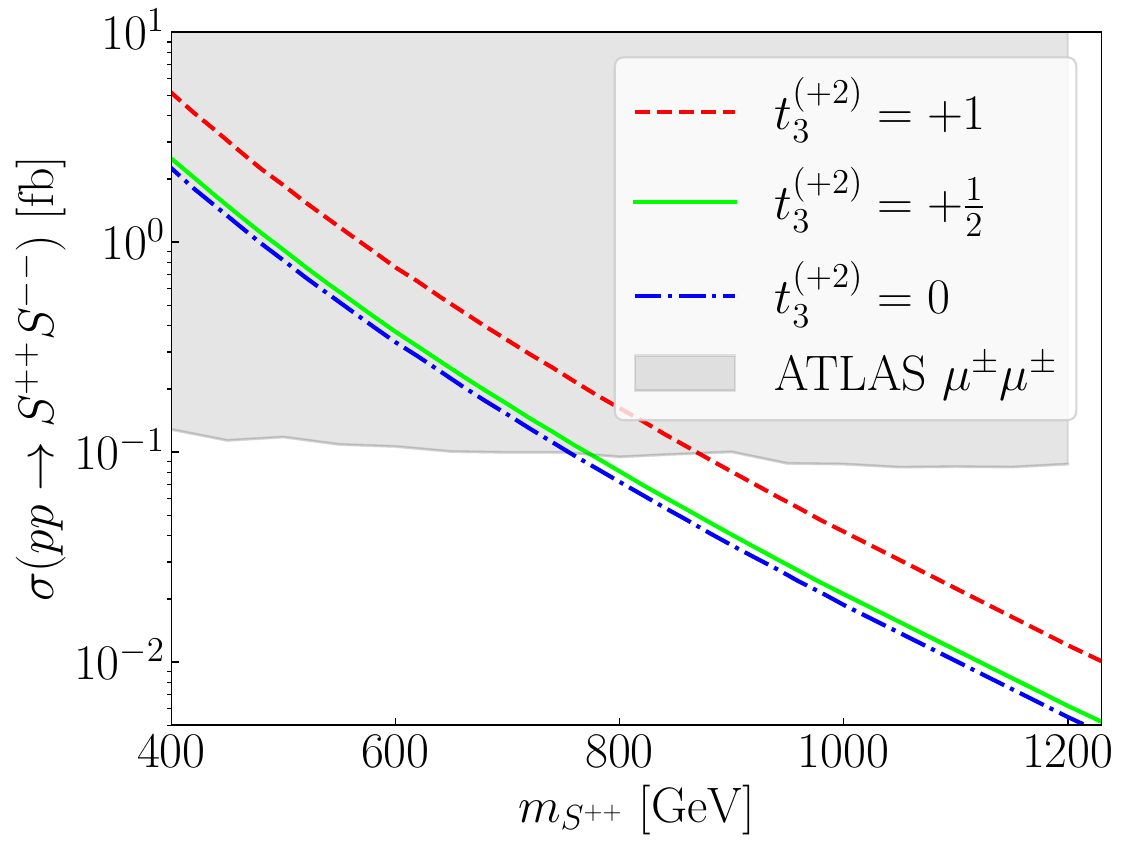}
    \includegraphics[width=0.42\textwidth,height=0.20\textheight]{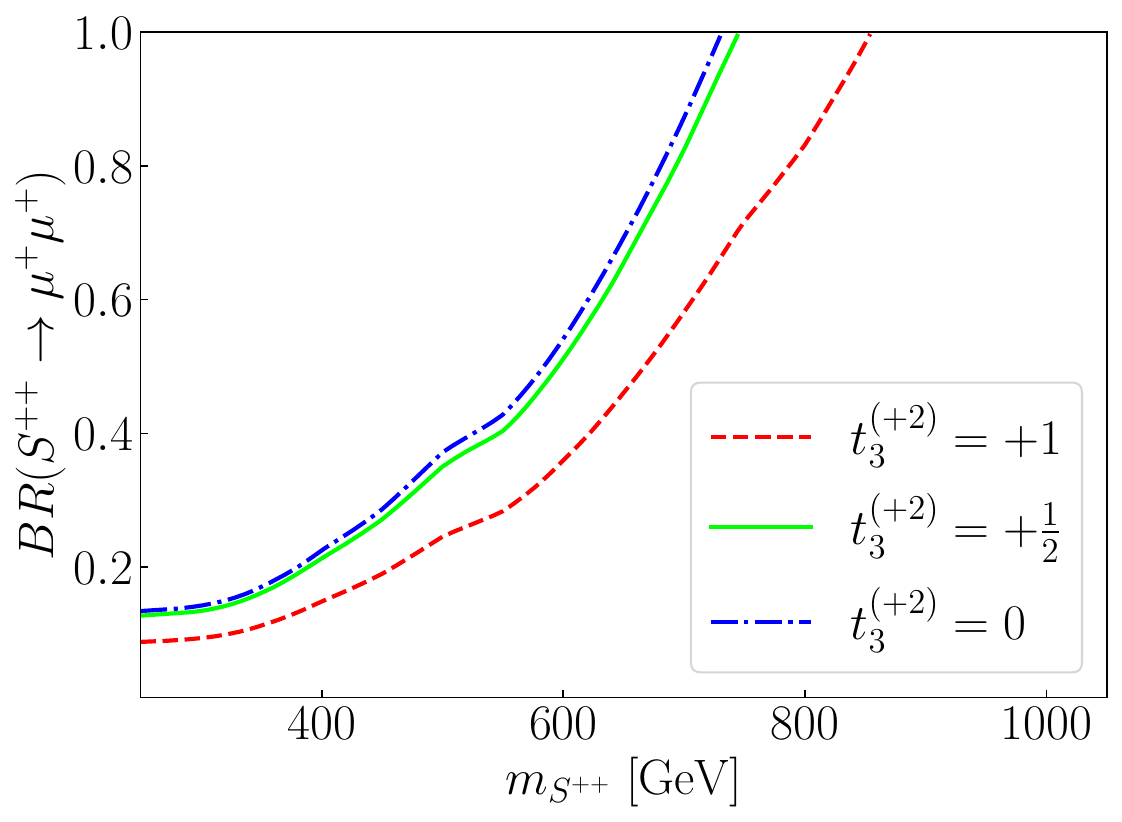}
    \includegraphics[width=0.42\textwidth,height=0.20\textheight]{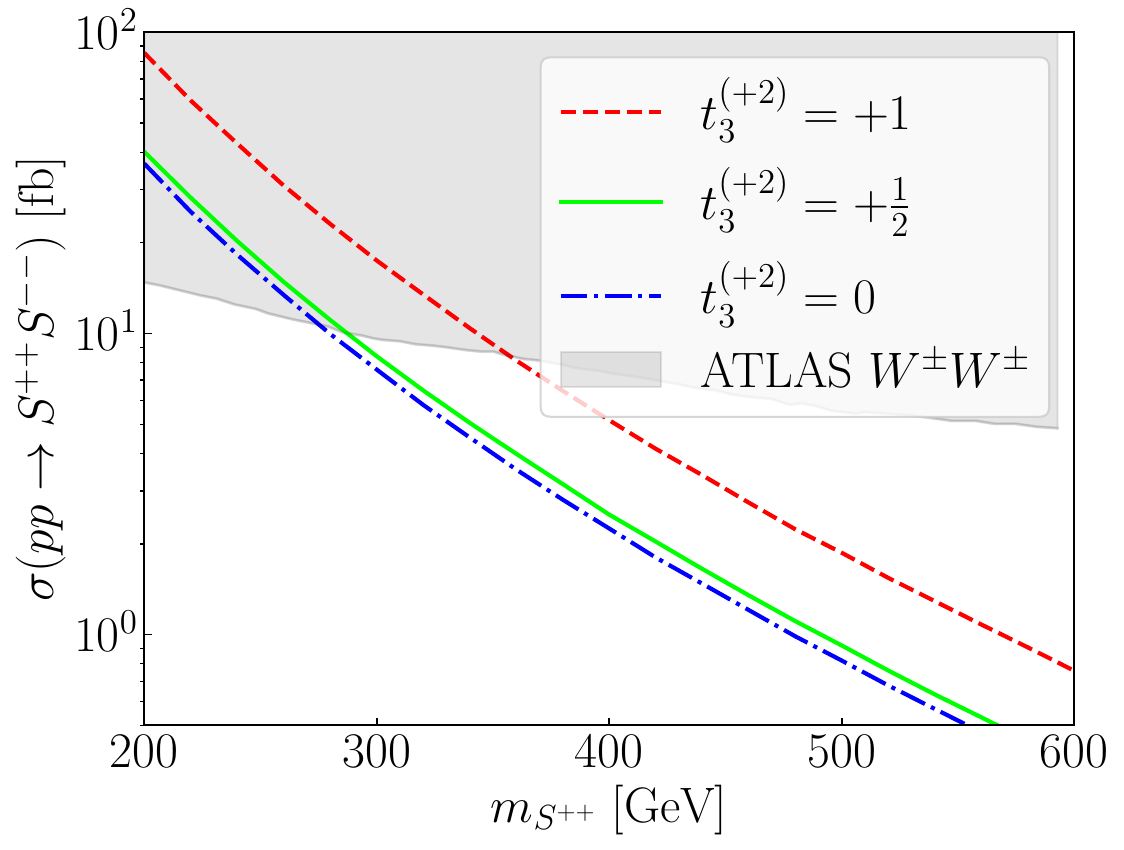}
    \includegraphics[width=0.42\textwidth,height=0.20\textheight]{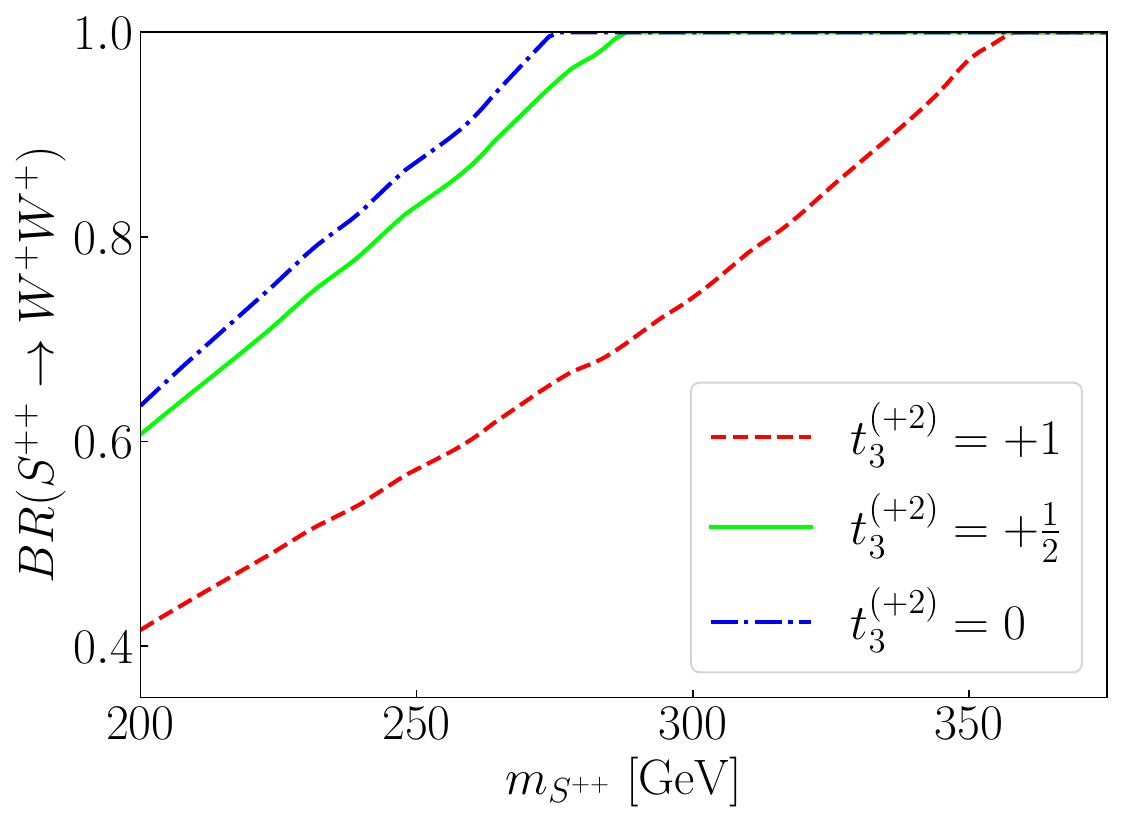}
    \caption{\small\textit{\textbf{Left:} The DY pair production cross-section $\sigma(pp\to S^{++}S^{--})$ at $\sqrt{s}=13$ TeV for doubly charged scalars as a function of its mass $m_{S^{++}}$ for different isospin quantum numbers, $\ttwo=0$ (blue dash-dotted), $+\frac{1}{2}$ (green) and $+1$ (red dashed), are presented. The gray shaded regions in the left panels show the experimental upper bounds on the production cross-section from the ATLAS searches for doubly charged scalars in same sign dilepton final state, assuming 100\% $BR$ in the $\ell^+\ell^{\prime +}$ channels, where $\ell, \ell^\prime=e, \mu$ \cite{ATLAS:2017xqs}, and in the $W^+ W^+$ channel~\cite{ATLAS:2021jol}. \textbf{Right:} Limits mapped to the $BR(S^{++} \rightarrow \ell^+\ell^{\prime +})$ (top three on the right panel) and $BR(S^{++} \rightarrow W^+W^{+})$ (bottom right) as a function of $m_{S^{++}}$ for different $\ttwo$.}}
    \label{fig:same-sign-dilepton}
\end{figure}
In the left panels of Fig.\,\ref{fig:same-sign-dilepton} we present the bounds on the DY production cross-section of $S^{++}$ as a function of mass, assuming 100\% $BR$ in the $S^{++}\to \ell^+\ell^{\prime +}$, where $\ell,\ell^\prime =e,\mu$, and the $S^{++} \to W^+ W^+$ decay channels, respectively. Note that we have used the ATLAS Run 2 data at the 36 fb$^{-1}$ luminosity for the dilepton channels~\cite{ATLAS:2017xqs} instead of the more updated analysis with 139 fb$^{-1}$ luminosity \cite{ATLAS:2022pbd}, since the latter assumes a specific relation among the $BR$s of the dilepton channels.
Irrespective of the $\ttwo$ value, the bounds from the dilepton final states (even with less luminosity) are comparatively stronger than the diboson channels. On the right panels of Fig.\,\ref{fig:same-sign-dilepton}, we reinterpret the bounds in the $BR$ vs. $m_{S^{Q^+}}$ plane, employing the Eq.\,\eqref{eq:bounds_FF}. The limits derived for the doubly charged scalars with $\ttwo=1$ and $0$ match with those of $H_L^{++}$ and $H_R^{++}$ from the LRSM as reported in the ATLAS analysis~\cite{ATLAS:2017xqs}.

The cross-section for $\ttwo=+1$ is considerably larger compared to $\ttwo=0$, since the interference between the $Z$ and photon-mediated processes is constructive for the former case, whereas it is destructive in the latter. For $\ttwo=+\frac{1}{2}$, although the interference is constructive, the $S^{++}S^{--}Z$ coupling is an order of magnitude smaller compared to the $\ttwo=+1$ case, leading to a smaller production cross-section. We again emphasize that the bounds on the $BR$s are independent of specific constructions, for example the constraints for $\ttwo=1$ is applicable equally to the HTM and the GM model. Similarly, the $H_R^{++}$ in LRSM and the doubly charged scalar from the Zee-Babu model, both arising from $\ttwo=0$, will have same constraints.

For the sake of completeness, we also discuss the search for long-lived doubly charged scalars. For example, the doubly charged scalar arising from a doublet with $Y=\frac{3}{2}$~\cite{Enomoto:2021fql}, or in the Higgs septet model~\cite{Hisano:2013sn} can not decay into a pair of SM particles, rendering them good candidates for being long-lived. We use results from the CDF collaboration \cite{CDF:2005hfo} to place $t_3^{(+2)}$ dependent bounds on the $m_{S^{++}}$. Thus, even in the absence of any visible decay modes, we can still have some non-trivial bounds on the doubly charged scalar mass from the CDF searches.

\paragraph{Singly charged scalars:}
\label{sec:limit_SQ1}

\begin{figure}[t!]
    \centering
    \includegraphics[width=0.485\textwidth]{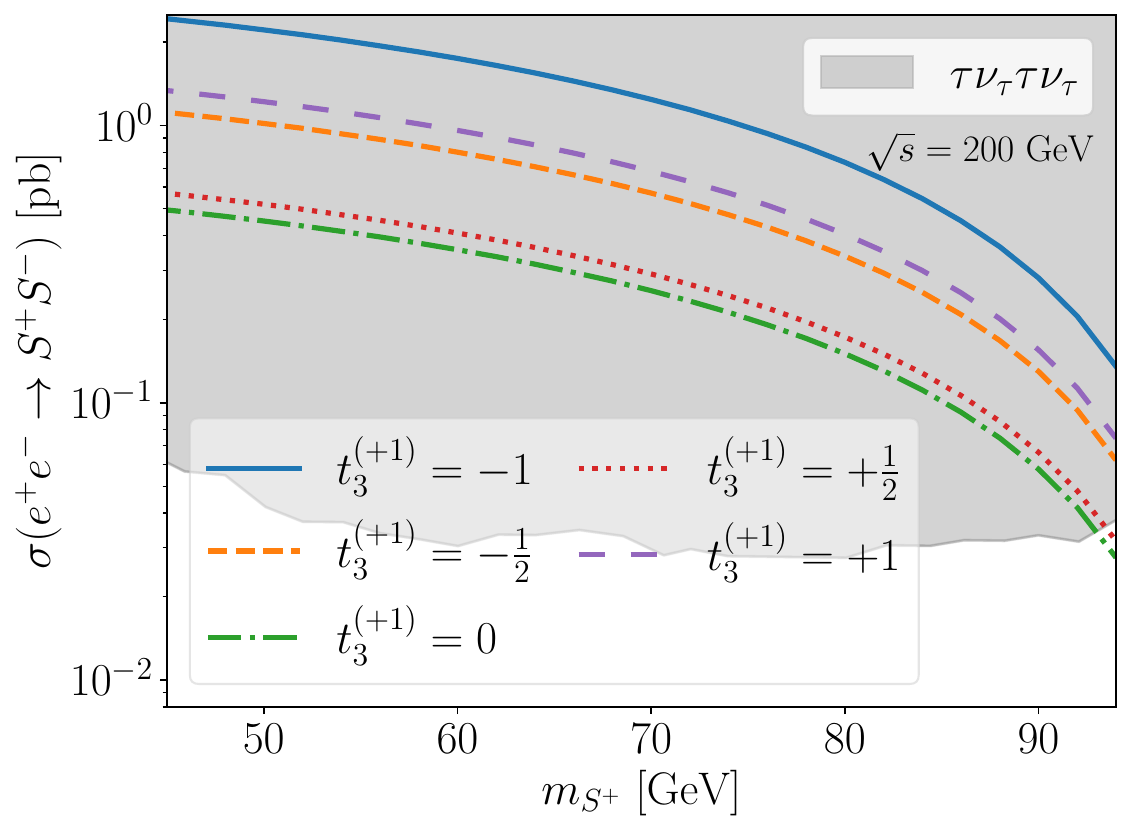}
    \includegraphics[width=0.485\textwidth]{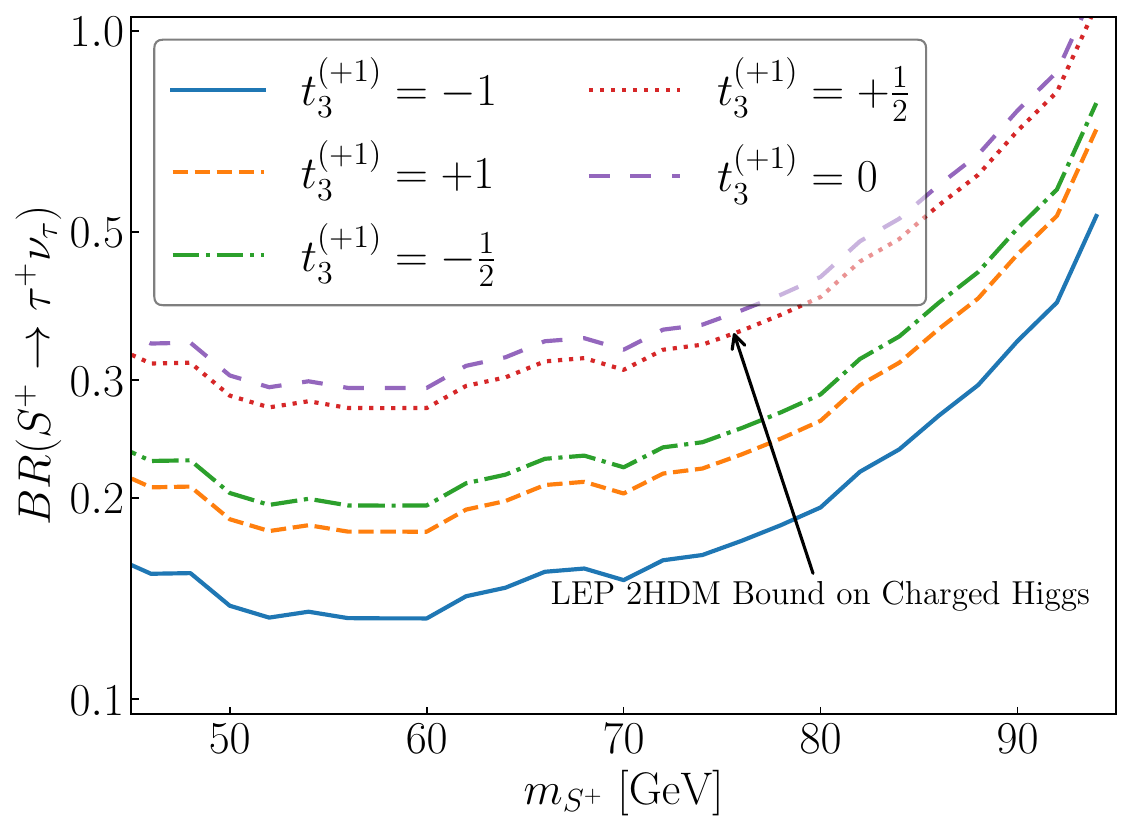}
    \caption{\small\textit{Extension of limits on the $BR(S^+ \to \tau^+ \nu)$ from the LEP data~\cite{ALEPH:2013htx} to different $\tone$ values within the range $[-1,+1]$. Notably, the LEP analysis begins at $M_{S^+}\gtrsim 45$~GeV, to prevent a large contribution to the $Z$-boson decay width due to the presence of the charged scalar, and is limited to $M_{S^+}\lesssim 95$~GeV due to the maximum operational energy of the LEP. 
    }}
    \label{fig:Brtaunu-mass-LEP}
\end{figure}

Next, we turn our attention to the DY pair production of singly-charged scalars. The singly charged scalars can decay into dilepton, diquark and diboson final states:
$$
S^+ \to \ell^+\nu, q_1 \bar{q}_2, W^+ Z, W^+ \gamma\,.
$$
First, let us examine the LEP data\cite{ALEPH:2013htx}, which focused primarily on the $S^+ \to \tau^+ \nu$ decay channel. It is important to note that the LEP analysis specifically assumes that the $S^+$ originates from a 2HDM, {\it i.e.}, the $\tone=+\frac{1}{2}$ component of a doublet, as previously mentioned. This provides an opportunity to extend the model-specific bounds to more general, model-agnostic $\tQ$-dependent limits in a straightforward manner. In the right panel of Fig.~\ref{fig:Brtaunu-mass-LEP}, we translate the limits on $\sigma(e^+e^-\to S^+S^-)$ from Ref.~\cite{ALEPH:2013htx}, shown in the left panel of the same figure, into $\tone$-dependent bounds on $BR(S^+\to \tau^+\nu)$, using Eq.\,\eqref{eq:bounds_FF}. Notably, our results for $\tone =+\frac{1}{2}$ are in excellent agreement with the corresponding bounds from the LEP analysis\cite{ALEPH:2013htx}, thereby validating our approach. Furthermore, we demonstrate how these bounds vary for other values of $\tone$.

\begin{figure}[t!]
    \centering
\includegraphics[width=0.485\textwidth]{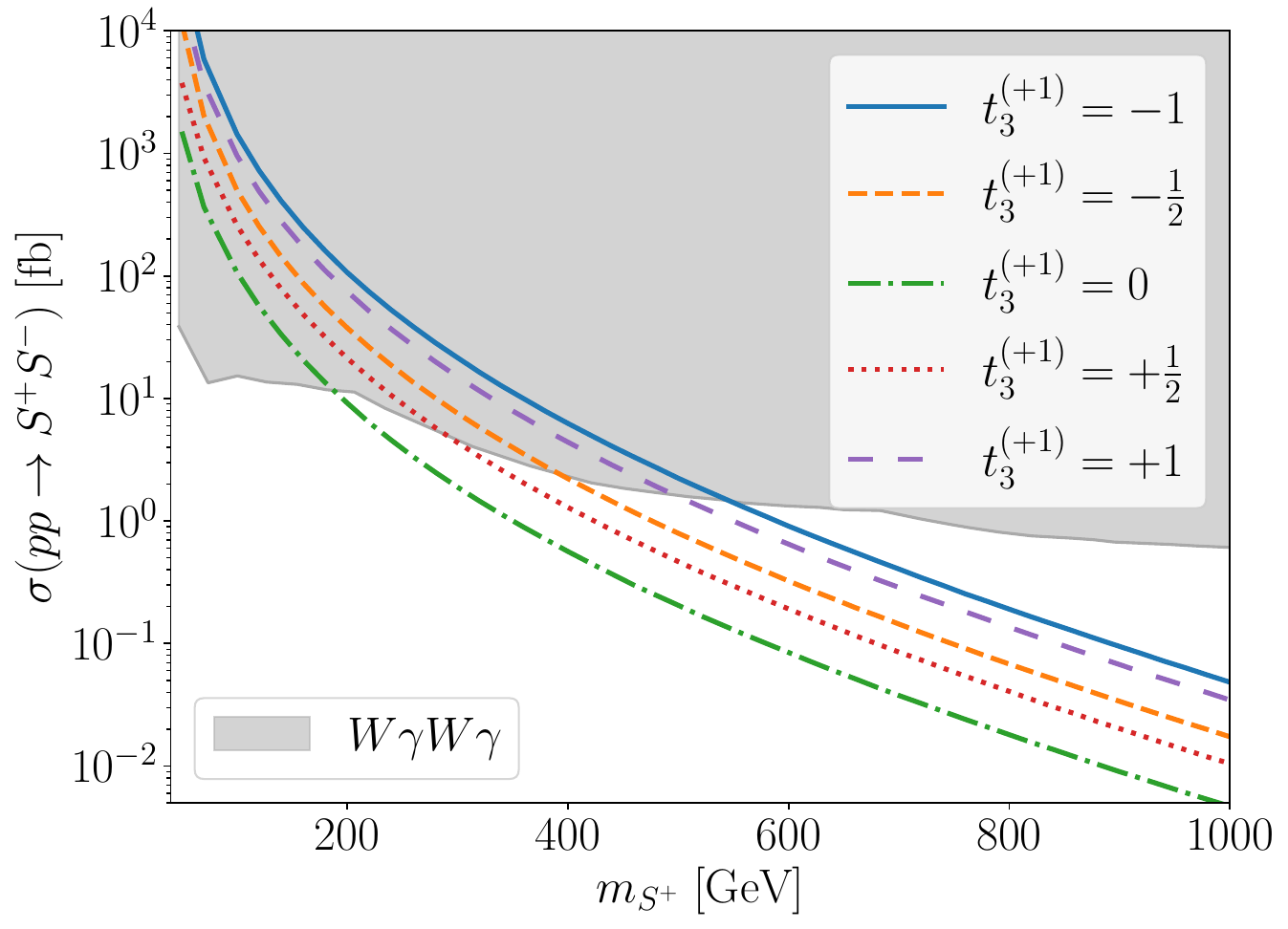}
\includegraphics[width=0.485\textwidth]{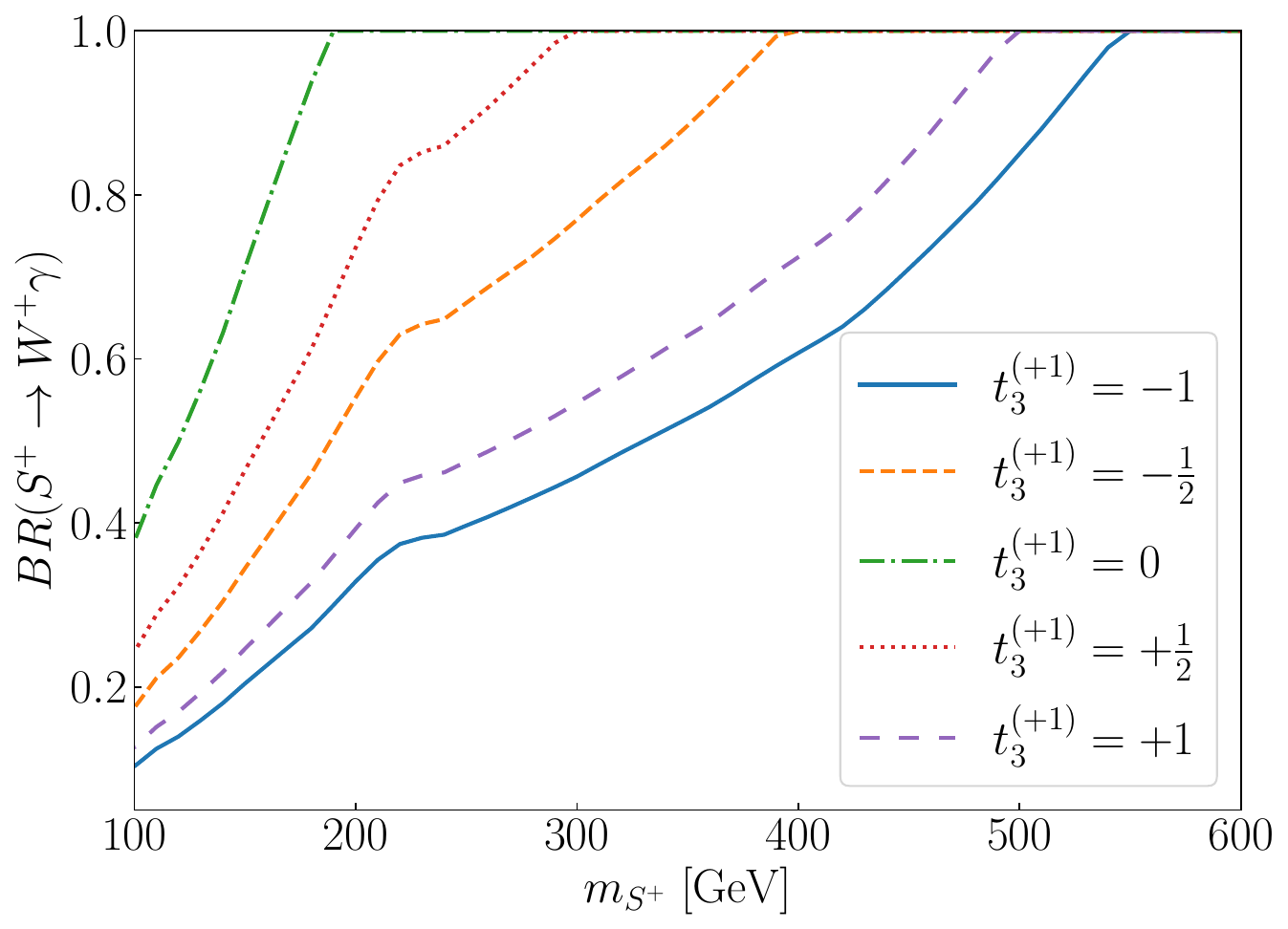}
\includegraphics[width=0.485\textwidth]{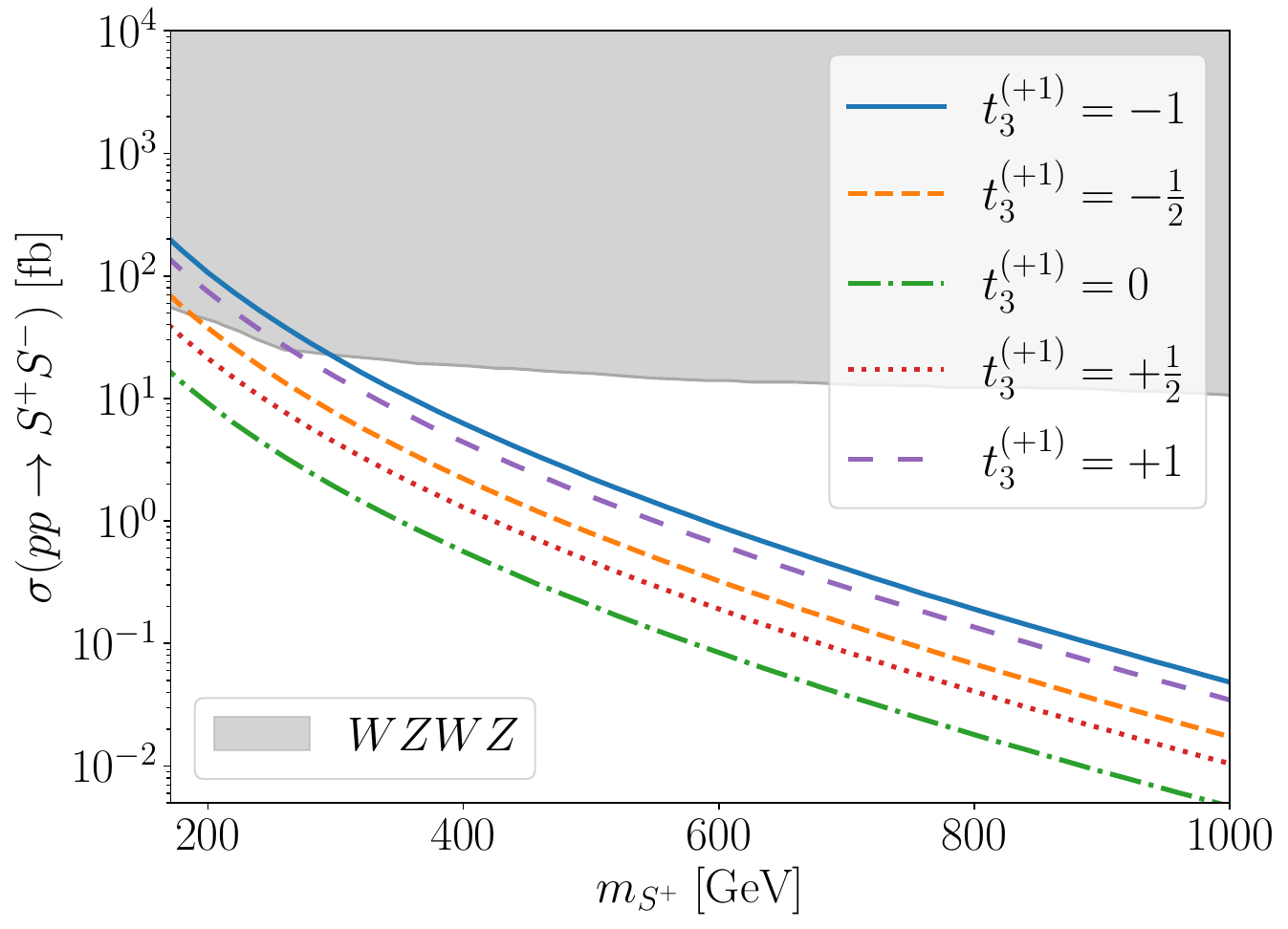}
   \includegraphics[width=0.485\textwidth]{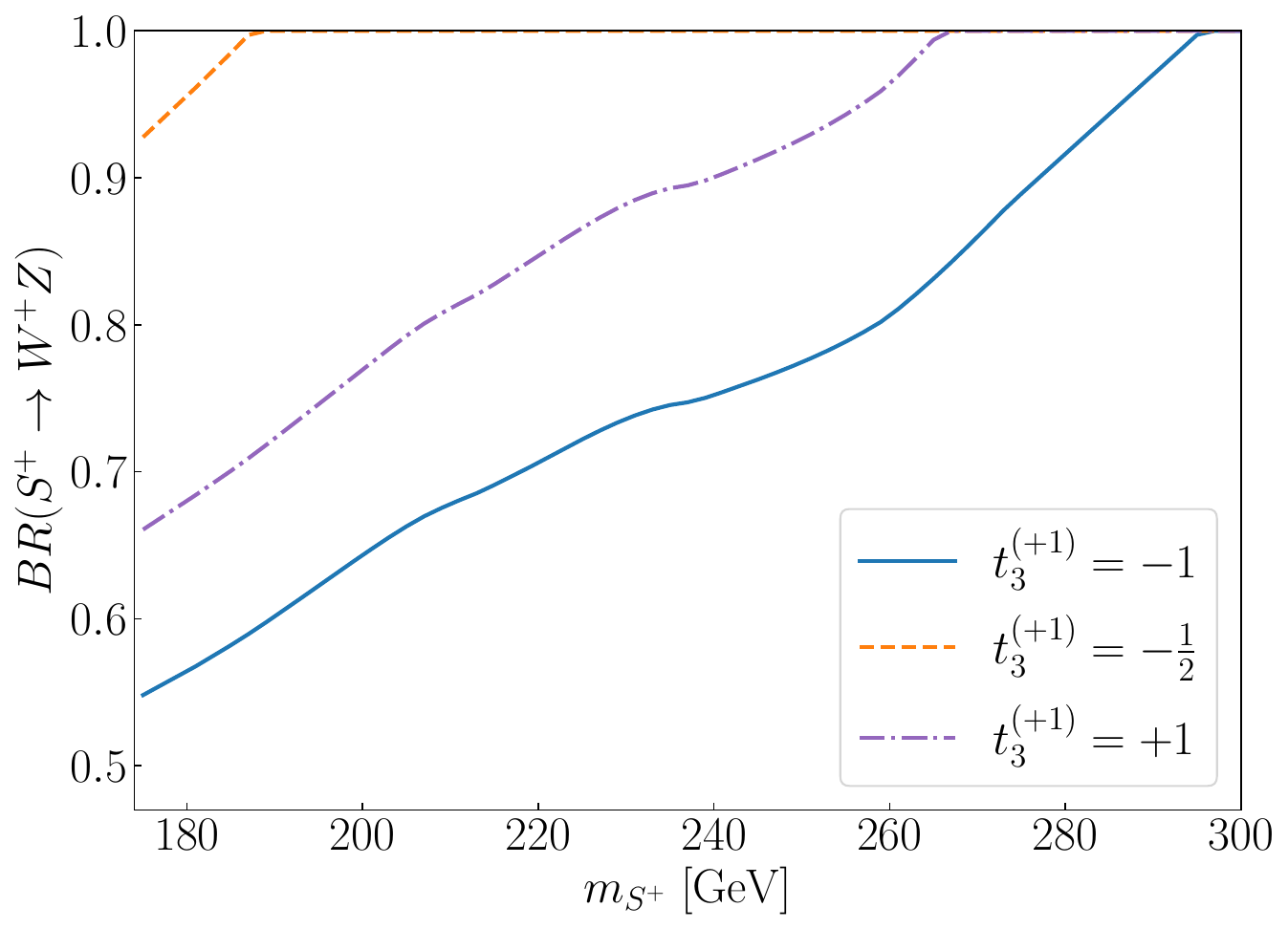}
    \caption{\textit{\small\textbf{Left:} The DY pair production cross-sections $\sigma(pp\to S^{+}S^{-})$ at $\sqrt{s}=13$ TeV for singly charged scalars as a function of their mass $m_{S^{+}}$ for $\tone\in [-1,1]$ are presented. The gray shaded regions show the upper bounds on the production cross-section, obtained from the phenomenological recast in~\cite{Cacciapaglia:2022bax}, assuming that $S^+$ decays into $W\gamma$ (top), $WZ$ (bottom) with 100\% BR, respectively.
    \textbf{Right:} The limits on the $BR(S^{+} \rightarrow W^+\gamma)$ (top), and $BR(S^{+} \rightarrow W^+Z)$ (bottom) as a function of $m_{S^+}$ for different isospin. Note that the current bounds for the $WZ$ channel are not sensitive to the isopsin quantum numbers $\tone=0$ and $+\frac{1}{2}$.}}
    \label{fig:singly-charged-crosssec-mass-Wg-Wz}
\end{figure}

Currently there is no dedicated search for singly charged scalar via DY production at the LHC. Thus, we shall rely on the bounds from the phenomenological analysis \cite{Cacciapaglia:2022bax}, which recasts various BSM searches at the LHC that yield the same final state objects. The analysis in Ref. \cite{Cacciapaglia:2022bax} focused on the decays of $S^+$ into third generation quarks and diboson final states.\footnote{The results from the LHC Run 2 (with 36 fb$^{-1}$ luminosity) is not yet sensitive to place any meaningful constraints on the decay channel $S^+ \to t\bar b$ for $\tone\in[-1,1]$.} The left panels of Fig.\,\ref{fig:singly-charged-crosssec-mass-Wg-Wz} show the bounds on the production cross-section as a function of mass, of a singly charged Higgs decaying into $W^+ \gamma$ (top) and $W^+ Z$ (bottom) final states with 100\% $BR$. In the right panels of Fig.\,\ref{fig:singly-charged-crosssec-mass-Wg-Wz}, we translate these bounds into the $BR$s of the $W^+\gamma$ (top) and $W^+ Z$ (bottom) channels as functions of the mass of the charged scalar. 

The $W\gamma$ decay channel in particular, while typically generated radiatively in specific models, can exhibit significant $BR$s if the charged scalar is fermiophobic~\cite{Logan:2018wtm}. For instance, the $H_5^\pm$ in the GM model, which are members of the fiveplet under the custodial $SU(2)$, does not couple to the SM fermions. Instead, the effective $H_5^\pm W^\mp \gamma$ vertex arises through one-loop contributions from the gauge bosons and other scalars in the model. Notably, the $H_5^\pm W^\mp Z$ vertex is suppressed in this model if the triplet VEV $(v_t)$ is much smaller compared to the doublet VEV ($v_d$). Consequently, $H_5^+$ in the GM model exhibits a significant $BR$ into $W^+\gamma$ over a wide range of parameter space. For instance, if $v_t/v_d \approx \order(10^{-4})$ or less, the $H_5^\pm$ decays exclusively in the $W\gamma$ channel~\cite{Logan:2018wtm}, justifying its role as a potential search channel. Apart from the GM model, charged pNGBs from the composite Higgs model based on the $SU(5)/SO(5)$ coset also exhibit large $BR$s into both $W^+ \gamma$ and $W^+ Z$ channels, where the necessary interaction is generated by the Wess-Zumino-Witten term~\cite{Agugliaro:2018vsu,Banerjee:2022izw}. 

From the Fig.~\ref{fig:singly-charged-crosssec-mass-Wg-Wz} it is clear that the strongest bound arises when both the charged Higgs bosons decay into $W\gamma$ channel. This is a final state containing two photons which constitutes a relatively cleaner signature.  
We highlight that the current constraints on the $S^+ \to W^+ \gamma$ channel are sensitive to branching ratios as low as 40\% for a charged scalar with a mass of 300 GeV. Thus, this particular channel opens a promising avenue for conducting a dedicated search for the singly charged scalar in the Run 3 and future versions of LHC.

\paragraph{Isospin dependent limits:}
\label{sec:isospin}

\begin{figure}[t!]
    \centering
    \includegraphics[width=0.485\textwidth]{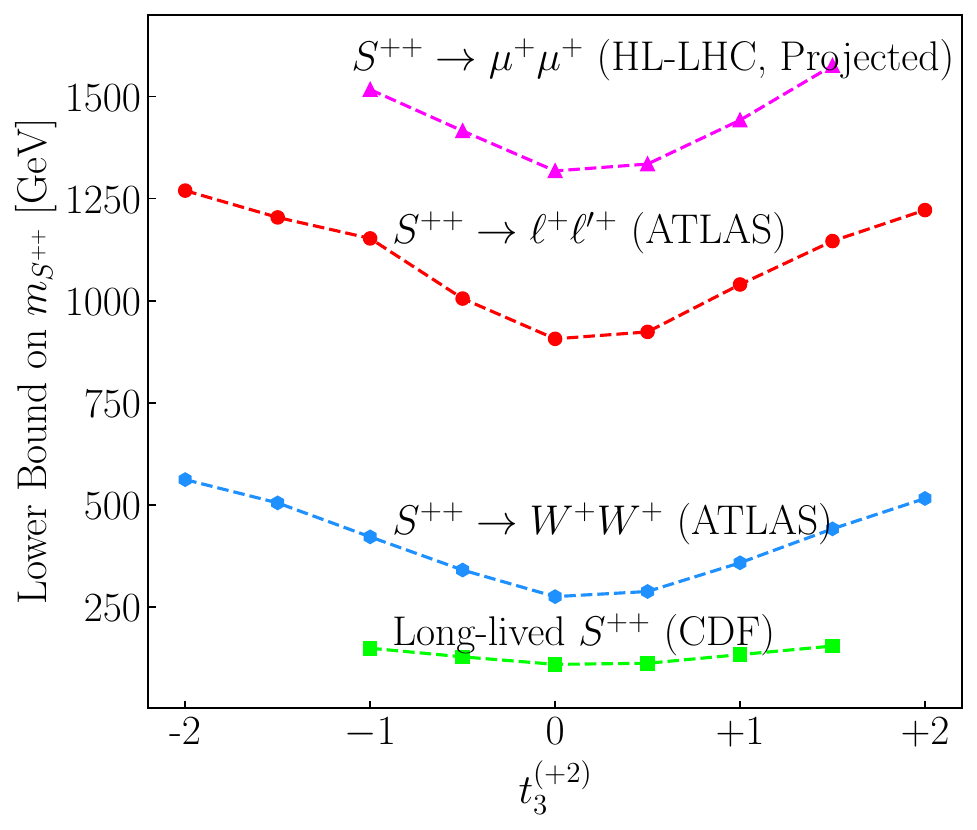}
    \includegraphics[width=0.485\textwidth]{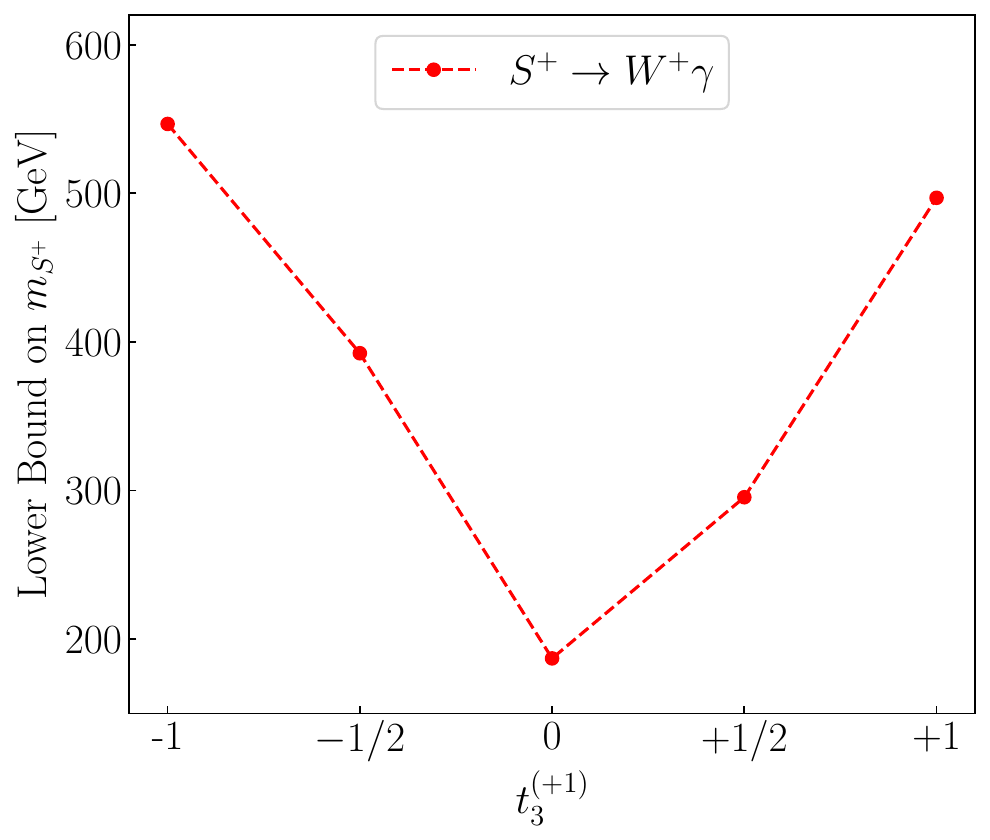}
    \caption{\small\textit{A comparison of lower limits on the masses of doubly (left) and singly (right) charged scalars arising from different $\tQ$ components of an isospin multiplet. For the doubly charged scalars the ATLAS constraints from the  $WW$ (blue) \cite{ATLAS:2021jol}, and $\ell \, \ell^\prime$ (red) \cite{ATLAS:2022pbd} decays are presented, assuming $100\%~BR$ in the corresponding decay channel. Additionally, for long-lived doubly charged scalars, limits from the CDF (green) \cite{CDF:2005hfo} are also included. The projection of constraints from the $\mu^+\mu^+$ channel at the HL-LHC \cite{CMS-PAS-FTR-22-006} is shown in magenta.  For singly charged scalars, only the bounds from $W \gamma$ channel\cite{Cacciapaglia:2022bax}, assuming 100\% $BR$, are shown. The lines joining the discrete $\tQ$ values are drawn only for clear visual representation.}}
    \label{fig:t3vsM}
\end{figure}
Fig.~\ref{fig:t3vsM} provides a clear overview of the relative importance of different search channels across various $\tQ$ values, helping to assess the effectiveness of each search channel for a given model.
For example, $H_5^{++}$ in the GM model originates from the $\ttwo=+1$ component of the complex triplet and decays predominantly in the $W^+ W^+$ final state. As shown in the left panel
of Fig.~\ref{fig:t3vsM}, this allows for a lower bound of $320$~GeV to be placed on the mass of $H_5^{\pm\pm}$
($m_5$) in the GM model. 
In the Zee-Babu model, when the doubly charged scalar arising from $\ttwo=0$ is the lightest BSM scalar, it predominantly decays into $\ell^+ \ell^{\prime +}$ channels, where $\ell,\ell^\prime = e, \mu,\tau$. The lower bound on the $m_{S^{++}}$ in this case can be as strong as $900$~GeV~\cite{ATLAS:2022pbd}.
However, projected limits at the HL-LHC \cite{CMS-PAS-FTR-22-006} indicate that doubly charged scalar having $\ttwo=+1$ and masses up to $1.5$~TeV can be excluded if it decays predominantly into the same sign dimuon channel.
In the left panel, we also show that $\ttwo$-dependent bounds of $\order(100~{\rm GeV})$ can be placed
on the mass of the doubly charged scalar, even if it is long-lived and does not have a visible
decay mode.
Similarly, the right panel of Fig.~\ref{fig:t3vsM} illustrates that significant limits on the mass of the singly-charged scalar can be derived when it decays predominantly into the $W\gamma$ channel.


So far, we have discussed charged scalars that have a well-defined $\tQ$ eigenvalue. 
However, there exist models where the charged scalars originate from a mixture of multiple
component fields with different $\tQ$ eigenvalues.
We wish to emphasize that our analysis can be easily extended to such mixed cases as well. To illustrate our point, let us consider singly charged scalars that arise from an admixture between $\tQ =0$ and $\tQ =+1$ components as follows:
\begin{equation} 
\begin{pmatrix}
  S_1^{\pm} \\ 
  S_2^{\pm} 
\end{pmatrix}
=
\begin{pmatrix}
  \cos\theta & \sin\theta\\ 
  -\sin\theta & \cos\theta
\end{pmatrix}
\begin{pmatrix}
  h_{(0)}^{\pm} \\ 
   h_{(1)}^{\pm}
\end{pmatrix}\,,
\end{equation}
where $S_1^{\pm}, S_2^{\pm}$ are physical charged scalars and $h_{(0)}^{\pm}, h_{(1)}^{\pm}$ are the unphysical components with $t_3^{(+1)} = 0$ and $+1$ respectively.
\begin{figure}[t!]
    \centering
    \includegraphics[width=0.58\textwidth,height=0.3\textheight]{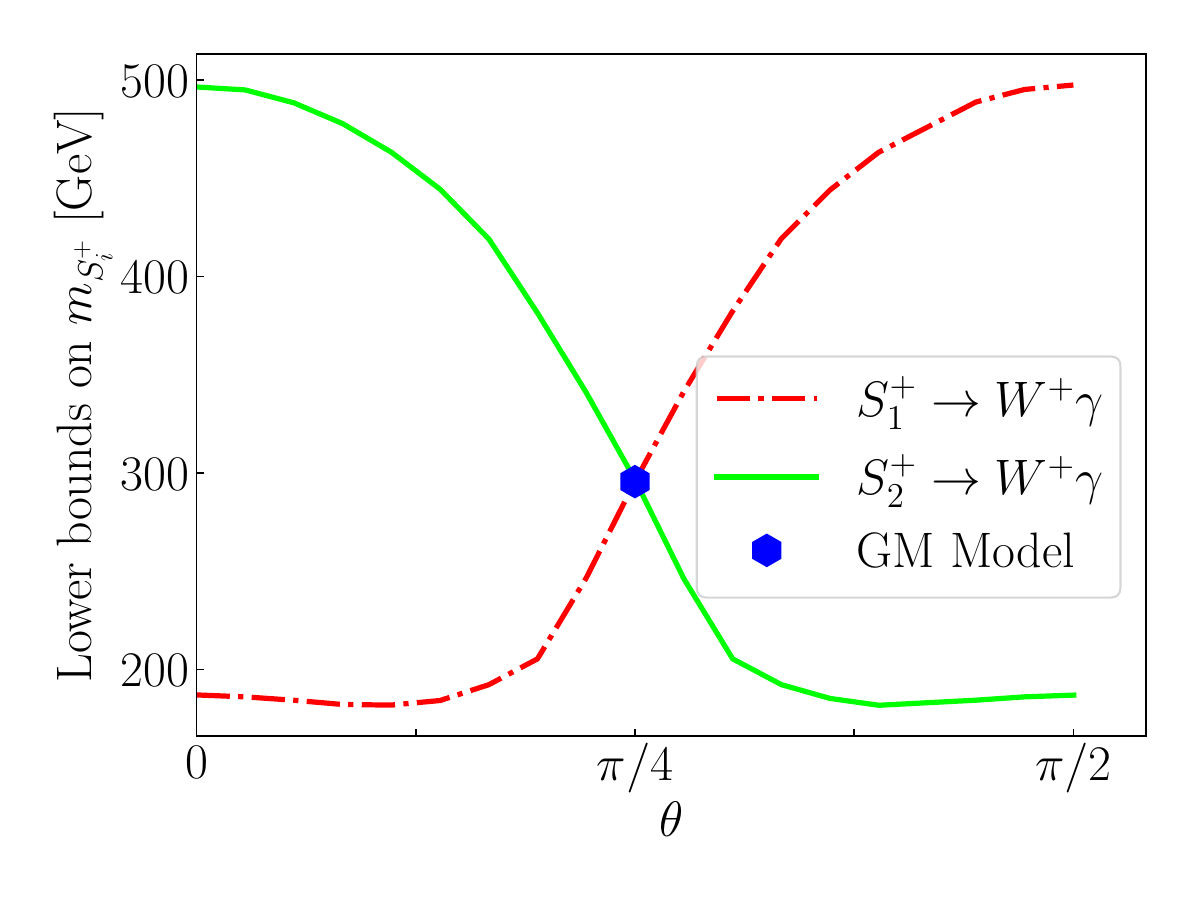}
    \caption{\small\textit{Lower limits on the masses of two physical charged scalar mass eigenstates $S_1^+$ (red) and $S_2^+$ (green) as a function of their mixing angle $\theta$, assuming 100\% $BR$ into the $W^+\gamma$ channel. The blue dot indicates the GM model, where the two charged scalars $H_3^+$ and $H_5^+$ has approximately equal amount of mixing, $\theta\simeq \pi/4$.}}
    \label{fig:Mixed-charged-scalar}
\end{figure}
In Fig.\,\ref{fig:Mixed-charged-scalar} we interpret the bounds on the decay channel $S^+\to W^+ \gamma$ as limits on the masses of the charged scalars as a function of the mixing angle, assuming 100\% $BR$. To exemplify, we specifically show the GM model where $S_{1}^{\pm}$, and $S_{2}^{\pm}$ can be identified as $H_3^\pm$ and $H_5^\pm$, respectively, while the mixing angle $\theta= \pi/4$. For $\theta = 0$, the limits on $S_1^\pm$ and $S_2^\pm$ align with those of a charged scalar having $t_3^{(+1)} = 0$ and $+1$, respectively. 
In contrast, for the GM model, where the mixing angle is $\theta = \pi/4$, the limits on the charged Higgs mass fall between those of the two unmixed scenarios, approximately at 300 GeV. Similar analysis can be performed for other mixed cases, such as the Higgs septet model and the Zee model. In the Higgs septet model, mixing occurs between the charged scalars from the  $t_3^{(+1)} = -1$ and $+3$  components, while in the Zee model, the physical charged scalars are mixture of the  $t_3^{(+1)} = 0$ and $+\frac{1}{2}$ components, as indicated in the Table~\ref{tab:my_label}. However, unlike the GM model the mixing angle remains a free parameter in these cases. 
\paragraph{Example of model specific limits (Higgs Triplet Model):}
\label{BR_limits_model}

Before closing the article, we demonstrate that the generic bounds obtained above can be easily translated into model-specific constraints on the parameter space, by expressing the branching ratios in terms of model parameters. As an example, let us consider the limits on the doubly-charged Higgs and try to interpret it in terms of the parameters of the HTM, which contains both singly and doubly charged scalars, as mentioned earlier. This shows an example where the limits on the $BR$s from Fig.\,\ref{fig:same-sign-dilepton} are easily expressed in terms of the model parameters. The phenomenology of this model, by and large, is governed by the triplet VEV ($v_t$). An interesting feature of this model is that the doubly charged scalar, usually denoted by  $H^{\pm\pm}$, decays mostly into same-sign dilepton for $v_t \lesssim \order(10^{-4})$~GeV and into same-sign $W$-bosons for $v_t \gtrsim \order(10^{-4})$~GeV, spanning complementary range of the parameter space \cite{Melfo:2011nx,Ashanujjaman:2021txz}. 

\begin{figure}[t!]
    \centering
    \includegraphics[width=0.48\textwidth]{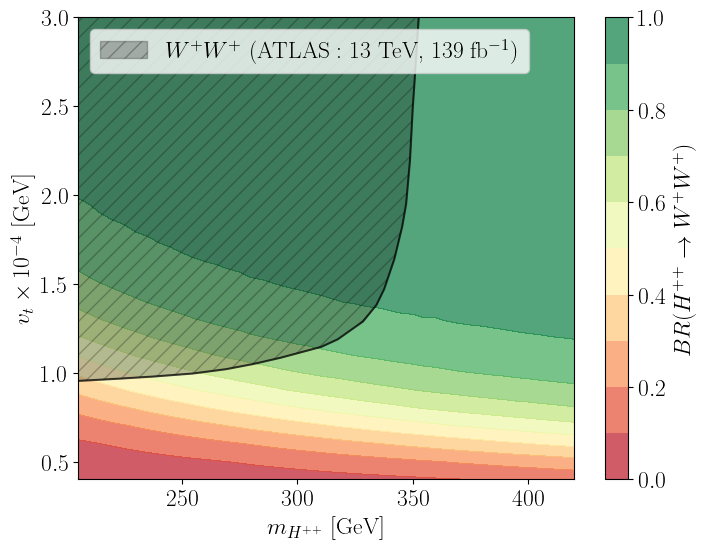}
\includegraphics[width=0.48\textwidth]{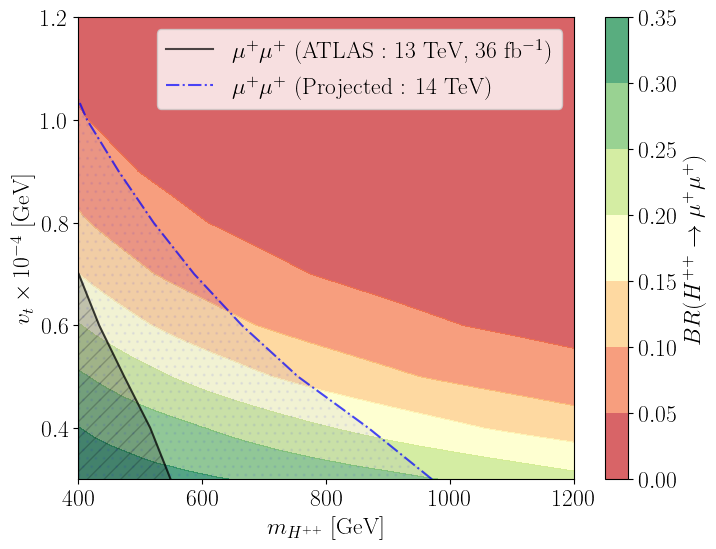}
    \caption{\small\textit{The branching ratios of $H^{++}$ to same-sign $W$-boson (left panel) and to same-sign dimuon (right panel) are shown in the $m_{H^{++}}$ vs. $v_t$ plane for the Higgs triplet model, assuming normal hierarchy for the neutrino masses. The hatched regions are excluded from the ATLAS searches \cite{ATLAS:2021jol,ATLAS:2017xqs} in the respective final states. The area left to the blue dash-dotted line in the right panel shows the projected exclusion at the HL-LHC from the dimuon channel \cite{CMS-PAS-FTR-22-006}.}}
    \label{fig:Type-II-seesaw}
\end{figure}

The $BR$s are obtained from the \texttt{UFO} files generated using the \texttt{FeynRules} implementation of the HTM \cite{Das:2016bir}, and the analytical expressions for the corresponding decay rates are given in \cite{Ashanujjaman:2021txz}. Fig.\,\ref{fig:Type-II-seesaw} shows the variation of the $BR(H^{++} \to W^+ W^+)$ (left panel) and the $BR(H^{++} \to \mu^+ \mu^{+})$ (right panel), in the $m_{H^{++}}$ vs. $v_t$ plane, assuming normal hierarchy of the neutrino masses. The hatched regions are excluded from the ATLAS observed upper limits on the $H^{++}H^{--}$  production cross-section times branching fraction of the respective search channels. 


\paragraph{Conclusions:}
\label{conc}

To summarize, this study presents a streamlined approach to analyze constraints on the charged scalars arising from the DY production at the LEP and the LHC. We have compiled a list of popular models featuring doubly and singly charged scalars with various isospins. We have analyzed the $W^{+}W^{+}, \ell^{+}\ell^{\prime +}$ decay channels for the doubly charged scalar, while $\tau^+ \nu$, $W^{+}Z$, and $W^{+}\gamma$ channels for singly charged scalars, and illustrated the corresponding bounds in the mass versus branching ratio plane. The results have been recasted into specific model-based example, such as the Higgs triplet model. The main takeaways from our analysis are described below:
\begin{itemize}
\item We provide an isospin dependent perspective for the DY pair production of charged scalars, offering a broader scope of interpreting the experimental data compared to model-specific analyses commonly found in the literature. By focusing on the generic properties, such as mass, electric charge, and weak isospin quantum numbers of the charged scalars, we present a model agnostic recipe that organizes the constraints on the branching ratios, applicable to a whole range of extended scalar sector models.  

\item Our plots provide a comprehensive visual catalog highlighting the relative merits of various search channels across different BSM scenarios. In particular, Figure~\ref{fig:t3vsM} showcases the significance of different search channels across various $\tQ$ values, demonstrating their effectiveness for probing specific models.

\item In the case of the GM model, the $H_5^{++}$ decays to $W\gamma$ with a 100\% branching ratio for $v_t/v < 10^{-4}$. A key observation made in this work is that the $H_5^{++} \to W^+\gamma$ decay mode imposes a bound $m_5 \geq 300$ GeV within the GM model from the current searches at the LHC. 

\item As an upshot of our analysis, the $W\gamma$ decay mode emerges as a promising search channel for singly-charged scalars in future experimental searches through DY pair production. 
\end{itemize}

Our approach in this work will be valuable for analyzing data from Run~3 of the LHC, its high-luminosity upgrade, and other proposed future colliders, offering a straightforward method for experiments to interpret charged scalar searches across a variety of theoretical models.  

\section*{Acknowledgments}

A.B. acknowledges support from the Department of Atomic Energy, Govt. of India. D.D. thanks the Anusandhan National Research Foundation (ANRF), India for financial support through grant no. CRG/2022/000565. The work of S.M. is supported by the IOE-IISc fellowship. S.P. acknowledges the Department of Science and Technology (DST) INSPIRE for the fellowship with the sanction no. DST/INSPIRE Fellowship/2021/IF210680. 

\bibliography{main}
\bibliographystyle{JHEP}
\end{document}